\documentclass[aps,prc,preprint,groupedaddress,showpacs]{revtex4}
\usepackage[dvips]{graphicx}
\usepackage{amsmath,amssymb}

\begin{document}
\title{Liquid-gas instability and superfluidity in nuclear matter}

\author{Masayuki Matsuzaki}
\email[]{matsuza@fukuoka-edu.ac.jp}
\affiliation{Department of Physics, Graduate School of Sciences, Kyushu University,
             Fukuoka 812-8581, Japan}

\affiliation{Department of Physics, Fukuoka University of Education, 
             Munakata, Fukuoka 811-4192, Japan}\thanks{permanent address}

\date{\today}

\begin{abstract}
  We study effects of the medium polarization on superfluidity in symmetric 
nuclear matter in a relativistic formalism.  An effect of the liquid-gas 
instability is emphasized.  We examine two types of decomposition of the 
nucleon propagator; the standard Feynman-density and the particle-hole-antiparticle 
ones. In both cases, the medium polarization effect is determined by a 
characteristic cancellation among the $\sigma$, the longitudinal $\omega$, and 
the $\sigma$-$\omega$ mixed polarizations.  The instability leads to increase of 
pairing gap.  Around the saturation density that is 
free from the instability the medium polarization enhances pairing gap in the 
former case whereas reduces in the latter.   At the lowest density that is also 
free from the instability the gap increases in both cases. 
\end{abstract}

\pacs{21.65.+f, 21.60.-n, 21.60.Jz}
\maketitle

\section{Introduction}

  Superfluidity in nuclear matter has long been studied mainly in pure neutron
matter from a view point of neutron-star physics such as cooling
rates and glitch phenomena~\cite{TT}.  In addition, that in
nuclear matter with finite $Z/N$ ratio is also becoming of interest as 
basic information for the structure theory of finite nuclei, since recent
development of RI-beam experiments makes it possible to study $N \simeq Z$
medium-heavy nuclei and neutron-rich light nuclei.

  At present, there are two ways to describe the fundamental properties such as
the saturation property of the finite-density nuclear many-body system; the
non-relativistic and the relativistic models. They are understood as describing
observed properties almost equally.  Among them, we here adopt the latter. 
The origin of quantum hadrodynamics (QHD) can be traced back to Duerr's relativistic
nuclear model~\cite{Du} that reformulated a non-relativistic field
theoretical model of Johnson and Teller~\cite{JT}.  Since Chin and Walecka
succeeded in reproducing the saturation property of symmetric nuclear matter
within the mean-field approximation~\cite{CW,Wa,Ch1},
QHD has not only been evolving beyond the mean-field approximation as a
many-body theory but also been enlarging its objects as a nuclear structure
model such as infinite matter $\rightarrow$ spherical $\rightarrow$ deformed
$\rightarrow$ rotating nuclei~\cite{Ri,SW1}.  These successes indicate that the
particle-hole channel interaction in QHD is realistic.  In contrast,  relativistic
nuclear structure calculations with pairing done so far have been using
particle-particle channel interactions
borrowed from non-relativistic models and therefore the particle-particle
channel in QHD has not been studied fully even in infinite matter.  Aside from
practical successes, this situation is unsatisfactory theoretically.
Therefore, in this paper, we present an effort to derive an in-medium
particle-particle interaction that is consistent with the relativistic mean
field (RMF) although only infinite matter can be discussed at the present stage.

  Up to now, there have been a lot of non-relativistic studies of pairing
in nuclear matter.  As for the particle-particle interaction entering into
the gap equation, many authors adopted bare interactions whereas others
adopted renormalized ones such as $G$-matrices.
Although, in the medium, renormalized interactions would be used intuitively,
following reasons support the use of bare interactions: 
(1) The Green's function formalism leads to the sum of the irreducible
diagrams~\cite{Mi,Ba1}, and its lowest order is the bare interaction.  
(2) The gap equation itself implies the
short-range correlation~\cite{CMS,Maru,TT,Ba1}.
In general, medium renormalizations are expected to enhance the gap by
reducing the short-range repulsion.  An interesting exception is the Gogny 
force. This is known to reproduce the pairing properties given by bare 
interactions at least at low densities~\cite{BE}. 
Anyway, as a next step, polarization diagrams should be considered. 
In the non-relativisitc framework, a lot of works have been done to study 
medium polarization effects on superfluidity in neutron 
matter~\cite{Cl,Ch2,AWP,WAP,Sc,He,SPR,Sh,SFB}.  All of them concluded that the 
medium polarization reduces pairing gap significantly.  A recent ab initio 
calculation~\cite{Fa}, however, claimed that the polarization effect is weak. 
Studies of polarization effects on symmetric nuclear matter have just 
begun recently. 
Reference~\cite{LSS} reported that the gap {\it increases} substantially. 
At the low density limit, Ref.~\cite{He} discussed that in the Fermi systems 
with 4 species the medium polarization enhances the gap. 

  Symmetric matter was also studied in the relativistic framework~\cite{CZL,BCF}. 
Both of these references reported that the vacuum polarization reduces the gap 
while the latter reported that the medium polarization enhances the gap. 
References~\cite{SLS,BCF} discussed that the increase of the gap is 
related to the existence of an instability. This is known as the liquid-gas 
instability~\cite{FH1,FH2,LH}. 

  Thus, in this paper, we investigate the liquid-gas instability and its effects 
on superfluidity in symmetric nuclear matter. Results on pure neutron matter 
are also mentioned briefly. Preliminary results were reported in Ref.~\cite{MP}. 

\section{Liquid-gas instability in relativistic random phase approximation}

\begin{figure}[htbp]
  \includegraphics[width=7cm]{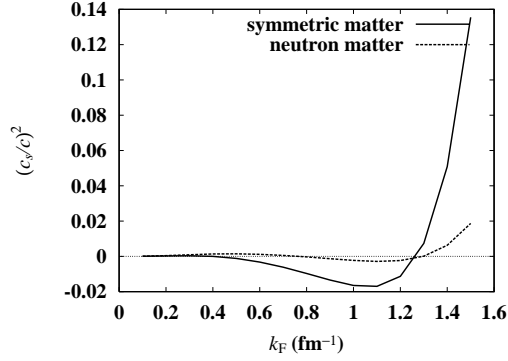}
 \caption{Squared ratios of the sound velocity to the light velocity as a function of 
the Fermi momentum. \label{fig1}}
\end{figure}

  We begin with the ordinary $\sigma$-$\omega$ model Lagrangian density,
\begin{gather}
\mathcal{L}=\bar\psi(i\gamma_\mu\partial^\mu-M)\psi
\notag\\
 +\frac{1}{2}(\partial_\mu\sigma)(\partial^\mu\sigma)
  -\frac{1}{2}m_\sigma^2\sigma^2
  -\frac{1}{4}\Omega_{\mu\nu}\Omega^{\mu\nu}
  +\frac{1}{2}m_\omega^2\omega_\mu\omega^\mu
\notag\\
 +g_\sigma\bar\psi\sigma\psi-g_\omega\bar\psi\gamma_\mu\omega^\mu\psi\, ,
\notag\\
\Omega_{\mu\nu}=\partial_\mu\omega_\nu-\partial_\nu\omega_\mu\, .
\end{gather}
Here $\psi$, $\sigma$, and $\omega$ are the nucleon, the $\sigma$ meson, and the 
$\omega$ meson fields, respectively, $M$, $m_\sigma$, and $m_\omega$ are their 
masses, and $g_\sigma$ and $g_\omega$ are the nucleon-meson coupling constants. 
The relativistic mean field approximation is carried out by replacing the 
meson fields in the coupled equations of motion by their expectation values as 
\begin{gather}
\sigma\rightarrow\langle\sigma\rangle=\sigma_0\, ,
\notag\\
\omega_\mu\rightarrow\langle\omega_\mu\rangle=\delta_{\mu 0}\omega_0\, .
\end{gather}
Then the nucleon effective mass (Dirac mass) equation is given by
\begin{equation}
\begin{split}
M^\ast
&=M-g_\sigma\sigma_0
\\
&=M-\frac{g_\sigma^2}{m_\sigma^2}\frac{\lambda}{\pi^2}
  \int_0^{k_\mathrm{F}} 
  \frac{M^\ast}{\sqrt{k^2+M^{\ast\,2}}}k^2dk\, ,
\end{split}
\label{effmass}
\end{equation}
where the isospin factor $\lambda$ = 2 and 1 indicate 
symmetric nuclear matter and pure neutron matter, respectively.
The Fermi momentum is related to the baryon density as 
\begin{equation}
\rho_\mathrm{B}=\frac{\lambda}{3\pi^2}k_\mathrm{F}^3\, .
\end{equation}
Properties of normal fluid, zero temperature matter of a given density is 
completely determined by the above effective mass equation (\ref{effmass}). 
The so-called saturation curve or the equation of state is given by the binding 
energy per nucleon, $E/A-M=\mathcal{E}/\rho_\mathrm{B}-M$, as a function of 
$\rho_\mathrm{B}$ or $k_\mathrm{F}$. This immediately gives pressure, 
\begin{equation}
P=\rho_\mathrm{B}^2\frac{\partial}{\partial\rho_\mathrm{B}}
\left(\frac{\mathcal{E}}{\rho_\mathrm{B}}\right)\, .
\end{equation}
Thermodynamic stability of the matter in the liquid phase is stated as 
$\partial P/\partial\rho_\mathrm{B}>0$. 
Since the ratio of the sound velocity $c_\mathrm{s}$ to the light velocity $c$ is given by 
\begin{equation}
\frac{c_\mathrm{s}}{c}=\sqrt{\frac{1}{M}\frac{\partial P}{\partial\rho_\mathrm{B}}}\, ,
\end{equation}
$\left(c_\mathrm{s}/c\right)^2<0$ in Fig.\ref{fig1} indicates the existence of a 
mechanical instability to the gas phase.  In this paper we adopt $M$ = 939 MeV, 
$m_\sigma$ = 550 MeV, $m_\omega$ = 783 MeV, $g_\sigma^2$ = 91.64, and 
$g_\omega^2$ = 136.2~\cite{SW2}.

\begin{figure}[htbp]
  \includegraphics[width=7cm]{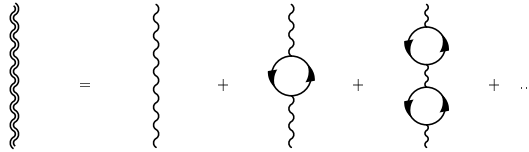}
 \caption{Feynman diagram representing the RPA meson propagator. \label{fig2}}
\end{figure}

  Quantum mechanically, stability of a state is determined by the second variation of the 
energy with respect to the fields~\cite{FH1}.  This is equivalent to the random phase 
approximation (RPA). The RPA in the present model in which the nucleon-nucleon 
interaction is mediated by mesons is formulated by calculating the meson propagators 
that couple to the particle-hole and particle-antiparticle 
polarizations~\cite{Ch1,FH2,LH}. The Dyson equation that determines the RPA propagator 
$D$ is given by 
\begin{equation}
D=D_0+D_0\Pi D
\label{dyson}
\end{equation}
(pictorially given in Fig.\ref{fig2}), where the lowest order propagator, 
\begin{equation}
D_0
=\left(
\begin{array}{@{\,}cc@{\,}}
D^\mathrm{S}_0 & 0 \\
0 & D^\mathrm{V}_{0 \mu\nu}
\end{array}
\right)\, ,
\end{equation}
and the polarization insertion,
\begin{equation}
\Pi
=\left(
\begin{array}{@{\,}cc@{\,}}
\Pi^\mathrm{S}     & \Pi^\mathrm{M}_\nu \\
\Pi^\mathrm{M}_\mu & \Pi^\mathrm{V}_{\mu\nu}
\end{array}
\right)\, ,
\end{equation}
are given by 5$\times$5 matrices. Their components are specified as
\begin{gather}
D^\mathrm{S}_0(q)=\frac{1}{q^2-m_\sigma^2+i\epsilon}\, ,
\notag \\
D^\mathrm{V}_{0 \mu\nu}(q)=\left(g_{\mu\nu}-\frac{q_\mu q_\nu}{m_\omega^2}\right)
                           D^\mathrm{V}_0(q)\,, \quad
D^\mathrm{V}_0(q)=\frac{-1}{q^2-m_\omega^2+i\epsilon}\, ,
\end{gather}
and
\begin{gather}
\Pi^\mathrm{S}(q)=-ig_\sigma^2\int\frac{d^4k}{(2\pi)^4}\mathrm{Tr}
\big[G(k)G(k+q)\big]\, ,
\notag\\
\Pi^\mathrm{V}_{\mu\nu}(q)=-ig_\omega^2\int\frac{d^4k}{(2\pi)^4}\mathrm{Tr}
\big[\gamma_\mu G(k)\gamma_\nu G(k+q)\big]\, ,
\notag\\
\Pi^\mathrm{M}_\mu(q)=ig_\sigma g_\omega\int\frac{d^4k}{(2\pi)^4}\mathrm{Tr}
\big[\gamma_\mu G(k)G(k+q)\big]\, .
\label{pi}
\end{gather}
Here $G(k)$ stands for the nucleon propagator and the Tr symbol includes isospin. 
$\Pi^\mathrm{M}_\mu$ stands for the matter induced $\sigma$-$\omega$ mixed polarization; 
a $\sigma$ excites a particle-hole pair and then it decays into an $\omega$ and 
vice versa.  Since Eq.(\ref{dyson}) is formally solved as
\begin{equation}
D=\frac{1}{1-D_0\Pi}D_0\, ,
\end{equation}
zeros of the dielectric function, 
\begin{equation}
\epsilon=\det{(1-D_0\Pi)}
\end{equation}
determines collective excitations. 

  For the present purpose, investigating instability, only the real parts at zero 
energy transfer are necessary.  Therefore we set 
\begin{gather}
D_0^\mathrm{S}=\frac{-1}{|\mathbf{q}|^2+m_\sigma^2}\, ,
\notag \\
D_0^\mathrm{V}=\frac{1}{|\mathbf{q}|^2+m_\omega^2}\, ,
\end{gather}
and the second term in $D^\mathrm{V}_{0 \mu\nu}$ drops because of the baryon 
number conservation.  This conservation law also restricts non-vanishing components 
among $\Pi$; if we choose the coordinate system as $q=(q^0,0,0,|\mathbf{q}|)$, 
only $\Pi^\mathrm{L}\equiv\Pi^\mathrm{V}_{00}-\Pi^\mathrm{V}_{33}$ and 
$\Pi^\mathrm{T}\equiv\Pi^\mathrm{V}_{11}=\Pi^\mathrm{V}_{22}$ among 
$\Pi^\mathrm{V}_{\mu\nu}$ and $\Pi^\mathrm{0}\equiv\Pi^\mathrm{M}_0$ among
$\Pi^\mathrm{M}_\mu$ survive. Note that energy transfer $q^0$ is set to zero, 
that is the instantaneous approximation. After some permutations, $1-D_0\Pi$ 
becomes block diagonal, and consequently the dielectric function reduces to 
\begin{gather}
\epsilon=\epsilon_\mathrm{L}\epsilon_\mathrm{T}^2\, ,
\notag \\
\epsilon_\mathrm{L}=(1-D_0^\mathrm{S}\Pi^\mathrm{S})(1-D_0^\mathrm{V}\Pi^\mathrm{L})
-D_0^\mathrm{S}D_0^\mathrm{V}(\Pi^0)^2\, ,
\notag \\
\epsilon_\mathrm{T}=1+D_0^\mathrm{V}\Pi^\mathrm{T}\, .
\end{gather}
The transverse dielectric function 
$\epsilon_\mathrm{T}$ is always positive in the density region in which superfluidity 
is realized, whereas the longitudinal one $\epsilon_\mathrm{L}$ becomes negative at 
intermediate densities; this represents the liquid-gas instability. 

  In this work, we concentrate on the particle-hole polarization. We examine two ways 
to extract the particle-hole polarization; (1) the standard Feynman-density (FD) 
decomposition of $G(k)$ and (2) the particle-hole-antiparticle (pha) decomposition. 
From its definition, the nucleon propagator in the medium is given by 
\begin{equation}
\begin{split}
G(k)
&=\frac{1}{2E^\ast(k)}
\Big[(\gamma_\mu K^\mu+M^\ast)
\Big(\frac{1-\theta(k_\mathrm{F}-|\mathbf{k}|)}{k^0-E^\ast(k)+i\epsilon}
    +\frac{\theta(k_\mathrm{F}-|\mathbf{k}|)}{k^0-E^\ast(k)-i\epsilon}
\Big) \\
&-(\gamma_\mu \tilde K^\mu+M^\ast)
\frac{1}{k^0+E^\ast(k)-i\epsilon}
\Big] \\
&\equiv G_\mathrm{p}(k)+G_\mathrm{h}(k)+G_\mathrm{a}(k)\, ,
\end{split}
\end{equation}
where
\begin{gather}
K=\big(E^\ast(k),\mathbf{k}\big) ,\quad \tilde K=\big(-E^\ast(k),\mathbf{k}\big)\, ,
\notag \\
E^\ast(k)=\sqrt{\mathbf{k}^2+M^{\ast\,2}}\, .
\end{gather}
The first, the second, and the third terms represent the propagator of particle, 
hole, and antiparticle, respectively. By sorting them with respect to the Heaviside 
function, another form, 
\begin{equation}
\begin{split}
G(k)
&=(\gamma_\mu k^\mu+M^\ast)\Big(\frac{1}{k^2-M^{\ast\,2}+i\varepsilon}
+\frac{i\pi}{E^\ast(k)}\delta(k^0-E^\ast(k))\theta(k_\mathrm{F}-|\mathbf{k}|)
\Big) \\
&\equiv G_\mathrm{F}(k)+G_\mathrm{D}(k)\, ,
\end{split}
\end{equation}
is obtained.  The first and the second terms are called the Feynman and the density 
parts, respectively.  The former consists of the antiparticle propagation and a part 
of the particle propagation, while the latter consists of the hole propagation and 
the other part of the particle propagation.  In the standard FD decomposition, 
the density dependent $G_\mathrm{F}G_\mathrm{D}+G_\mathrm{D}G_\mathrm{F}$ part in 
Eq.(\ref{pi}) is regarded as the particle-hole polarization. Note that the 
$G_\mathrm{D}G_\mathrm{D}$ part is pure imaginary and therefore is not necessary for 
the present purpose.  In the pha decomposition, the 
$G_\mathrm{p}G_\mathrm{h}+G_\mathrm{h}G_\mathrm{p}$ part in Eq.(\ref{pi}) represents 
the particle-hole polarization and this directly corresponds to that in 
non-relativistic calculations~\cite{Na}. The concrete expressions of $\Pi$ are 
given in Appendix A. 

\begin{figure}[htbp]
  \includegraphics[width=8cm]{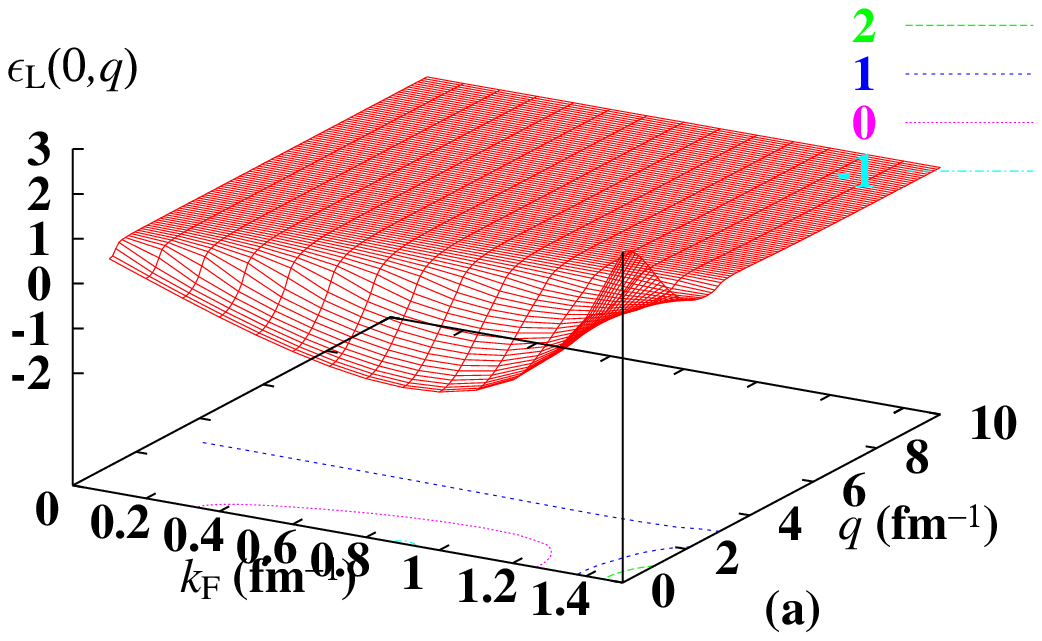}
  \includegraphics[width=8cm]{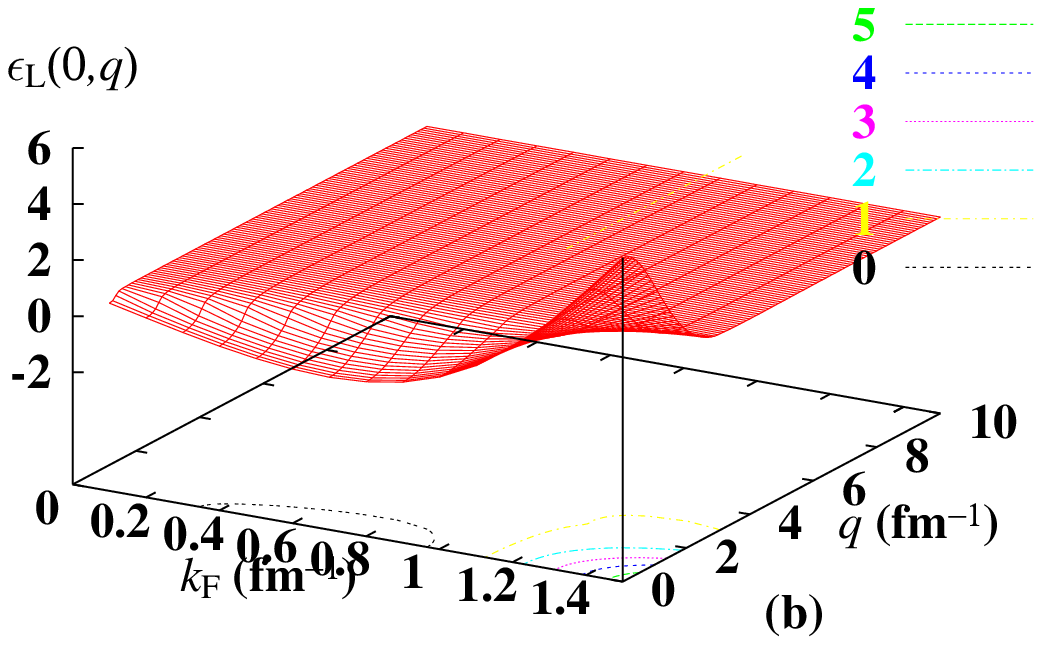}
 \caption{(Color online) Longitudinal dielectric functions for symmetric nuclear 
matter as a function of the Fermi momentum and the momentum transfer, (a) the FD 
case, (b) the pha case. \label{fig3}}
\end{figure}

 Figure~\ref{fig3} shows the longitudinal dielectric function $\epsilon_\mathrm{L}$ 
of the FD and the pha cases for symmetric nuclear matter.  At low momentum transfers, 
$\epsilon_\mathrm{L}$ becomes negative in both cases.  The density range of the 
instability coincides very well with Fig.~\ref{fig1} in the FD case.  This is 
consistent with Ref.~\cite{FH1} in which the density range of the instability hardly 
changes even after inclusion of the vacuum polarization.  In the pha case the 
instability region shrinks.  These results indicate that the liquid-gas instability 
occurred in a wide density range affects the pairing properties that are calculated 
by using the RPA meson propagator that exhibits instability. 

\section{Medium polarization effects on superfluidity}

  The concrete form of the RPA meson propagator is given by
\begin{equation}
D
=\frac{1}{1-D_0\Pi}D_0
=\left(
\begin{array}{@{\,}ccccc@{\,}}
\frac{(1-D_0^\mathrm{V}\Pi^\mathrm{L})D_0^\mathrm{S}}{\epsilon_\mathrm{L}} & \frac{D_0^\mathrm{S}D_0^\mathrm{V}\Pi^0}{\epsilon_\mathrm{L}} & 0 & 0 & 0 \\
\frac{D_0^\mathrm{V}D_0^\mathrm{S}\Pi^0}{\epsilon_\mathrm{L}} & \frac{(1-D_0^\mathrm{S}\Pi^\mathrm{S})D_0^\mathrm{V}}{\epsilon_\mathrm{L}} & 0 & 0 & 0 \\
0 & 0 & \frac{-D_0^\mathrm{V}}{\epsilon_\mathrm{T}} & 0 & 0 \\
0 & 0 & 0 & \frac{-D_0^\mathrm{V}}{\epsilon_\mathrm{T}} & 0 \\
0 & 0 & 0 & 0 & -D_0^\mathrm{V}
\end{array}
\right)\, .
\label{mesonprop}
\end{equation}
This gives the particle-particle channel interaction that determines 
superfluidity as
\begin{gather}
V_\mathrm{RPA}=\sum_{a,b}g_a\Gamma_a D^{a,b}g_b\Gamma_b\, ,
\notag \\
g_a=\left\{
\begin{array}{@{\,}r}
 g_\sigma \\
-g_\omega
\end{array}
\right. ,
\quad
\Gamma_a=\left\{
\begin{array}{@{\,}rl}
1          & :a=-1 \\
\gamma_\mu & :a=\mu
\end{array}
\right. .
\end{gather}
The antisymmetrized matrix element of this interaction for the $^1S_0$ pairing 
channel is given by 
\begin{equation}
\bar v(\mathbf{p},\mathbf{k})
=\langle\mathbf{p}s',\tilde{\mathbf{p}s'}\vert V_\mathrm{RPA}\vert
           \mathbf{k}s,\tilde{\mathbf{k}s}\rangle
   -\langle\mathbf{p}s',\tilde{\mathbf{p}s'}\vert V_\mathrm{RPA}\vert
           \tilde{\mathbf{k}s},\mathbf{k}s\rangle ,
\end{equation}
here the argument $|\mathbf{q}|$ of $D$ in $V_\mathrm{RPA}$ is specified as 
$\mathbf{q}=\mathbf{p}-\mathbf{k}$, and tildes represent time reversal. 
Since, by ignoring the coupling to the negative-energy states, the relativistic 
Numbu-Gor'kov formalism reduces to the usual gap equation~\cite{KR,MM}, 
\begin{equation}
\Delta(p)
=-\frac{1}{8\pi^2}
  \int_0^\infty \bar v(p,k)
         \frac{\Delta(k)}{\sqrt{(E_k-E_{k_{\rm F}})^2+\Delta^2(k)}}k^2dk\, ,
\label{gapeq}
\end{equation}
after an integration with respect to the angle between $\mathbf{p}$ and $\mathbf{k}$.
The effective mass equation (\ref{effmass}) is slightly 
modified when superfluidity sets in as
\begin{gather}
  M^\ast=M-\frac{g_\sigma^2}{m_\sigma^2}\frac{\lambda}{\pi^2}
  \int_0^\infty 
  \frac{M^\ast}{\sqrt{k^2+M^{\ast\,2}}}v_k^2k^2dk\, ,
\notag \\
v_k^2=\frac{1}{2}\Big(1-
\frac{E_k-E_{k_{\rm F}}}{\sqrt{(E_k-E_{k_{\rm F}})^2+\Delta^2(k)}}\Big)\, .
\end{gather}
Here $E_k=E^\ast(k)+g_\omega\langle\omega_0\rangle$.
Using these equations we calculate superfluidity in nuclear matter assuming that 
it is in the liquid phase even at the density range in which the liquid-gas 
instability would occur, as usual.  We set the upper bound of the integrations as 
20 fm$^{-1}$. 

  Here, some discussion about our choice of the interaction is in order. 
In the lowest order (the tree level; $D\rightarrow D_0$), $V_\mathrm{RPA}$ reduces 
to the one boson exchange (OBE) interaction $V_\mathrm{OBE}$ given by the RMF 
vertices (see Fig.\ref{fig2}). It has been well known that this OBE interaction gives 
unphysically large pairing gap~\cite{KR}.  Its reason can be traced back to the 
fact that the RMF vertices were tuned only below the Fermi momentum. In order to 
discuss the higher order (polarization) effects this OBE result must be improved 
beforehand.  We accomplish this by introducing a form factor that modulates the 
high momentum part of the interaction smoothly~\cite{MT1} as follows. Note that 
the authors of Ref.~\cite{BCF} adopted sudden momentum cutoffs so as to reproduce 
the virtual state in the $T$ matrix. We argued in Ref.~\cite{MT1} that the 
sudden cutoff distorts the shape of the short range pair wave function. 
See also Ref.~\cite{TM} for the sudden 
momentum cutoff that reproduces the results of a bare interaction in the pairing 
calculation.  From a general argument, the lowest order in the particle-particle 
channel interaction that gives pairing should be a bare interaction and the 
particle-hole channel interaction that determines the medium polarization should 
be an in-medium one in principle.  In the present investigation, however, we 
calculate both the tree and the bubble contributions on the same footing adopting 
an interaction of in-medium nature, which reproduces the results of a bare 
interaction in the tree level. That is, we regard, in a sense, the present RMF-OBE 
interaction with a form factor resembles the Gogny force in the non-relativistic 
pairing calculations. 

  In order to modulate the high momentum interaction that enters into the pairing 
calculation, we introduce a form factor, 
\begin{equation}
f(\mathbf{q}^2)=\frac{\Lambda^2}{\Lambda^2+\mathbf{q}^2}\, ,
\end{equation}
at each vertex.  This does not affect the mean field (Hartree) part with the momentum 
transfer $\mathbf{q} = 0$. 
The parameter $\Lambda$ is determined so as to minimize the difference 
in the pairing properties from the results of the RMF+Bonn calculation, 
that is, a hybrid calculation performed by adopting single particle states 
from the RMF model and the Bonn potential as the pairing interaction.  
Here we adopt the Bonn-B potential because this has a moderate property 
among the available (charge-independent) versions A, B, and C~\cite{Mach}. 
The pair wave function, 
\begin{gather}
\phi(k)=\frac{1}{2}\frac{\Delta(k)}{E_{\rm qp}(k)}\, , 
\notag \\
\,E_{\rm qp}(k)=\sqrt{(E_k-E_{k_{\rm F}})^2+\Delta^2(k)}\, ,
\label{pairwf}
\end{gather}
is related to the gap at the Fermi surface,
\begin{equation}
  \Delta(k_{\rm F})=-\frac{1}{4\pi^2}
  \int_0^\infty \bar v(k_{\rm F},k)\phi(k)k^2dk\, ,
\end{equation}
and its derivative determines the coherence length, 
\begin{equation}
\xi=\left(\frac{\int_0^\infty\vert\frac{d\phi}{dk}\vert^2k^2dk}
               {\int_0^\infty\vert\phi\vert^2k^2dk}\right)
      ^\frac{1}{2}\, ,
\end{equation}
which measures the spatial size of the Cooper pairs. These expressions 
indicate that $\Delta(k_{\rm F})$ and $\xi$ carry independent information, 
$\phi$ and $\frac{d\phi}{dk}$, respectively, in strongly-coupled systems, 
whereas they are intimately related to each other in weakly-coupled ones. 
Therefore we search for $\Lambda$ that minimizes
\begin{equation}
\chi^2
=\frac{1}{2N}\sum_{k_{\rm F}}
\left\{\left(\frac{\Delta(k_{\rm F})_{\rm RMF}-\Delta(k_{\rm F})_{\rm Bonn}}
                   {\Delta(k_{\rm F})_{\rm Bonn}}\right)^2
+\left(\frac{\xi_{\rm RMF}-\xi_{\rm Bonn}}
            {\xi_{\rm Bonn}}\right)^2
\right\}\, ,
\end{equation}
with $N = 11$ ($k_\mathrm{F} = 0.2, 0.3, ..., 1.2$ fm$^{-1}$).  The obtained 
value is $\Lambda = 7.26$ fm$^{-1}$~\cite{MT1}.  For consistency we included 
the form factor also in the polarization diagrams but it almost would not 
affect them because the polarization affects only low momenta (see 
following figures). 

\begin{figure}[htbp]
  \includegraphics[width=7cm]{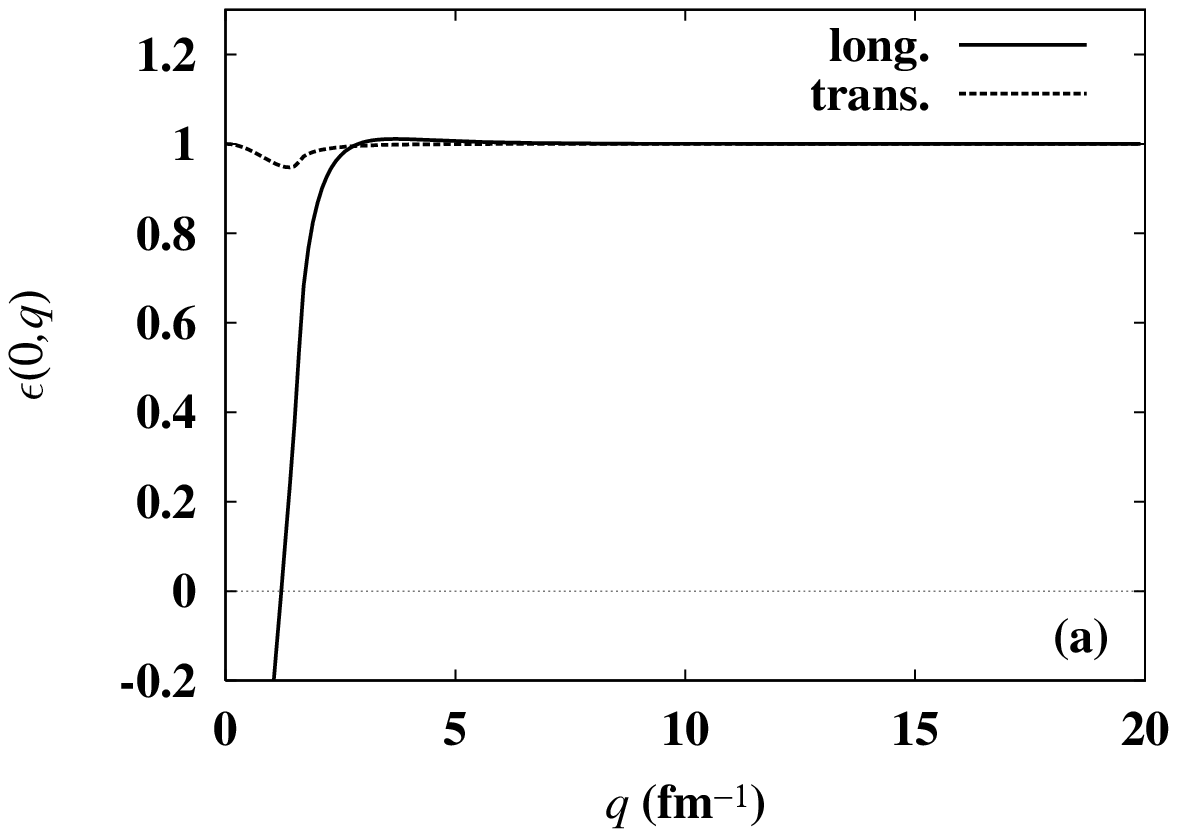}
  \includegraphics[width=7cm]{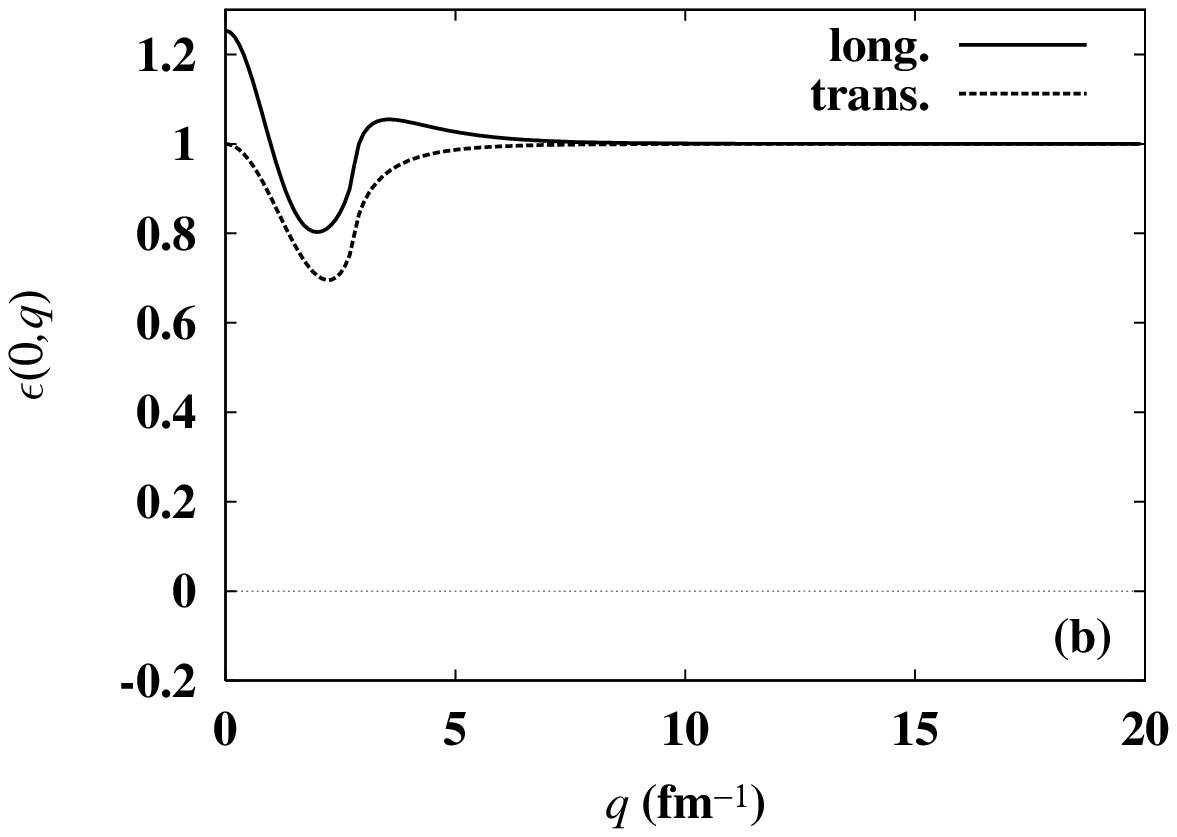}
 \caption{Cross sections of the dielectric functions for the FD case of symmetric 
nuclear matter as a function of the momentum transfer, (a) 
$k_\mathrm{F} = 0.8$ fm$^{-1}$, (b) $k_\mathrm{F} = 1.4$ fm$^{-1}$. \label{fig4}}
\end{figure}

\begin{figure}[htbp]
  \includegraphics[width=7cm]{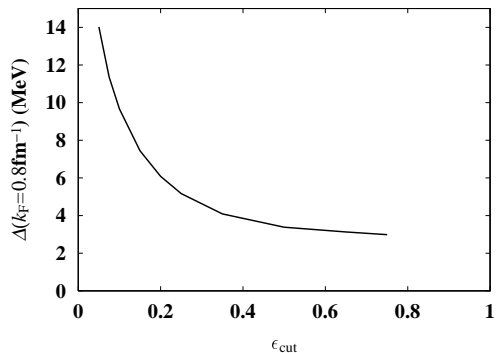}
 \caption{Cutoff parameter dependence of $\Delta(k_\mathrm{F})$ at 
$k_\mathrm{F} = 0.8$ fm$^{-1}$. \label{fig5}}
\end{figure}

\begin{figure}[htbp]
  \includegraphics[width=7cm]{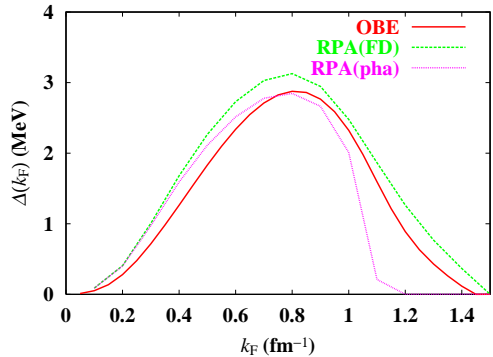}
 \caption{(Color online) Pairing gaps at the Fermi surface for symmetric 
nuclear matter as a function of the Fermi momentum. \label{fig6}}
\end{figure}

\begin{figure}[htbp]
  \includegraphics[width=7cm]{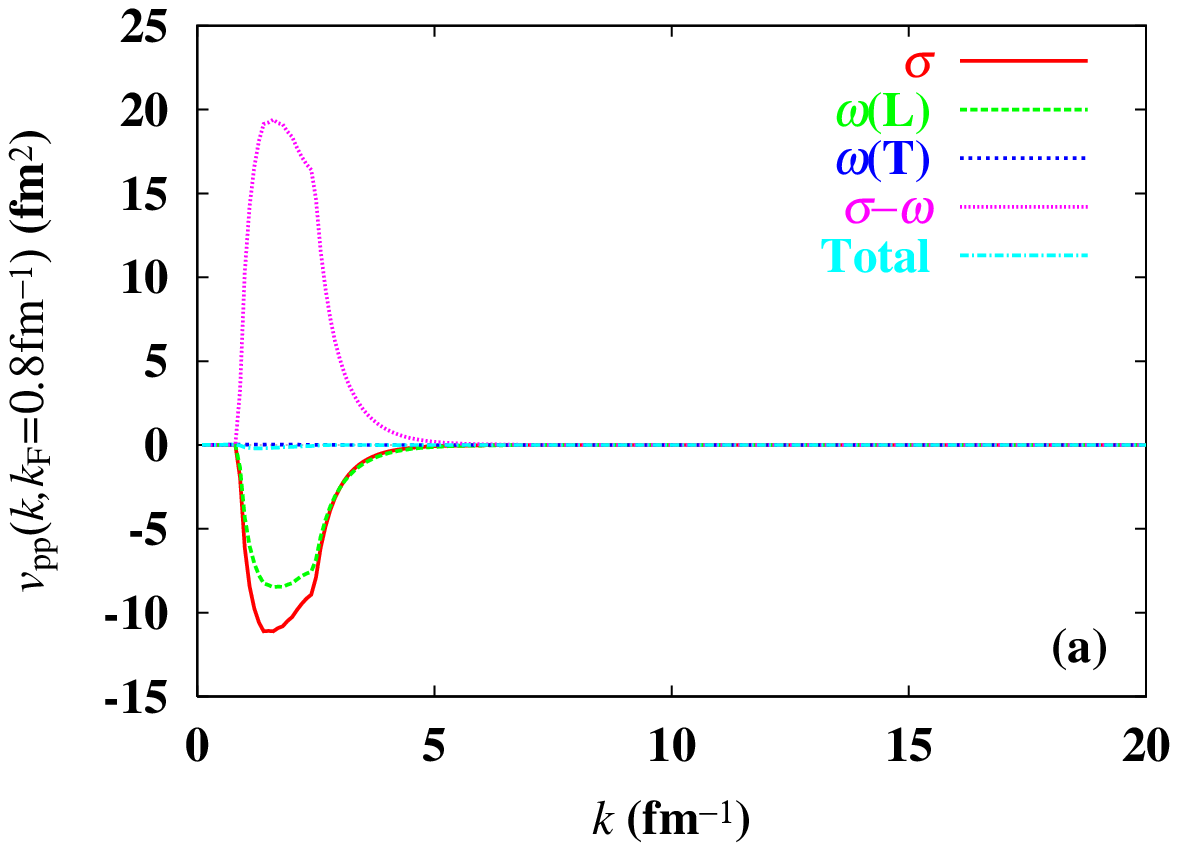}
  \includegraphics[width=7cm]{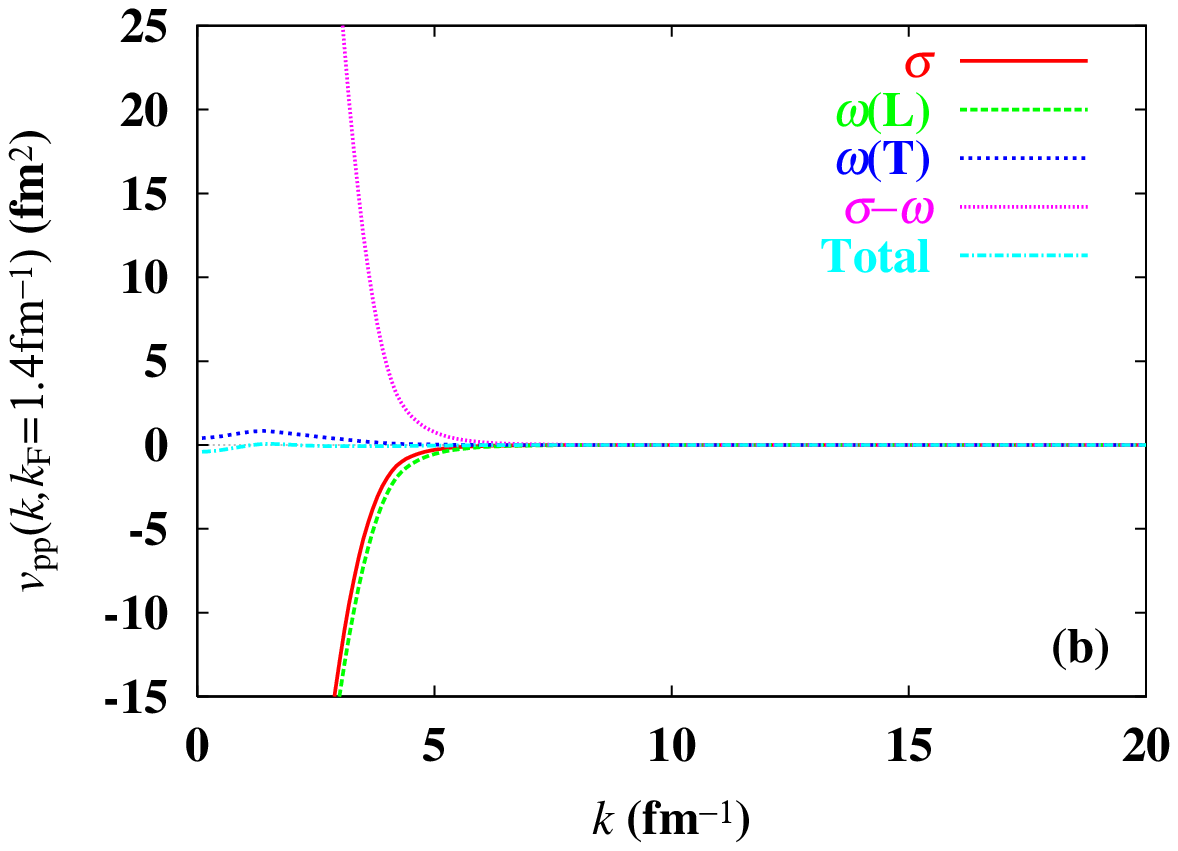}
 \caption{(Color online) Decomposition of the polarization interaction for the 
FD case of symmetric nuclear matter as a function of the momentum, (a) 
$k_\mathrm{F} = 0.8$ fm$^{-1}$, (b) $k_\mathrm{F} = 1.4$ fm$^{-1}$. \label{fig7}}
\end{figure}

  Equation (\ref{mesonprop}) indicates that $V_\mathrm{RPA}$ becomes 
ill-defined when the liquid-gas instability occurs.  This means that 
superfluidity in the liquid-gas coexistent phase rather than that in the 
liquid phase should be considered.  However, in the present calculation, we 
consider the case in which the system stays in the liquid phase also in the 
instability region (0.3 fm$^{-1} < k_\mathrm{F} <$ 1.3 fm$^{-1}$ in the FD 
case) as usual.  Therefore in that region only qualitative discussion is possible. 
Reference~\cite{SLS} also mentioned the existence of the instability. 
Note that discussion is quantitative in high and low density regions 
that are free from the instability. 
Figure~\ref{fig4} shows the cross sections of the longitudinal and the 
transverse dielectric functions.  Figure~\ref{fig4}(a) is for 
$k_\mathrm{F} =$ 0.8 fm$^{-1}$ and (b) is for $k_\mathrm{F} =$ 1.4 fm$^{-1}$. 
The former indicates that a low momentum cutoff is necessary to regularize the 
calculation for the instability region.  Thus, we introduce 
$\epsilon_\mathrm{cut}$ that operates to cut $|\mathbf{q}|$ for which 
$\epsilon_\mathrm{L/T}(0,|\mathbf{q}|)\le\epsilon_\mathrm{cut}$.  Since other 
parameters were determined at the saturation density, we chose 
$\epsilon_\mathrm{cut}$ = 0.65 that maintains the full variation of 
$\epsilon_\mathrm{L}$ and $\epsilon_\mathrm{T}$ around this density (see 
Fig.\ref{fig4}(b)).  This also serves to make the $k_\mathrm{F}$-dependence 
of $\Delta(k_\mathrm{F})$ smooth at the boundary of the instability region. 
Figure~\ref{fig5} reports the dependence of 
$\Delta(k_\mathrm{F}=0.8$ fm$^{-1})$ on $\epsilon_\mathrm{cut}$. 
Its dependence is moderate around the chosen value.  Figure~\ref{fig6} compares 
$\Delta(k_\mathrm{F})$.  This figure shows that the medium polarization 
increases the gap at all densities in the FD case.  This is common to the 
previous calculation~\cite{MP} in which the form factor to modulate the 
high momentum interaction was not introduced.  Batista et al.~\cite{BCF} 
who weakened the effect of the bubble diagram by their $x$ parameter 
obtained a similar result.  The contents of the polarization interaction, 
$V_\mathrm{RPA}-V_\mathrm{OBE}$, 
are decomposed in Fig.~\ref{fig7}.  (The $k$-dependence in Fig.~\ref{fig7}(a) 
is a result of $\epsilon_\mathrm{cut}$ for $q$.)  These figures represent a 
characteristic feature that the $\sigma$ polarization and the longitudinal 
$\omega$ polarization give strong attractions whereas the $\sigma$-$\omega$ 
mixed polarization gives a strong repulsion and they strongly cancel each 
other.  Remaining tiny attraction leads to the increase of $\Delta(k_\mathrm{F})$. 
From this viewpoint, the neglect of the $\sigma$-$\omega$ mixed 
polarization in Ref.~\cite{CZL} would cause an imbalance.  The transverse 
$\omega$ polarization that represents the spin density fluctuation is slightly 
repulsive, but the repulsion in the momentum region in which 
$\Delta(k) < 0$ (that is predominantly determined by $V_\mathrm{OBE}$) 
also increases $\Delta(k_\mathrm{F})$ because of the structure of the gap 
equation (\ref{gapeq}).  


\begin{figure}[htbp]
  \includegraphics[width=7cm]{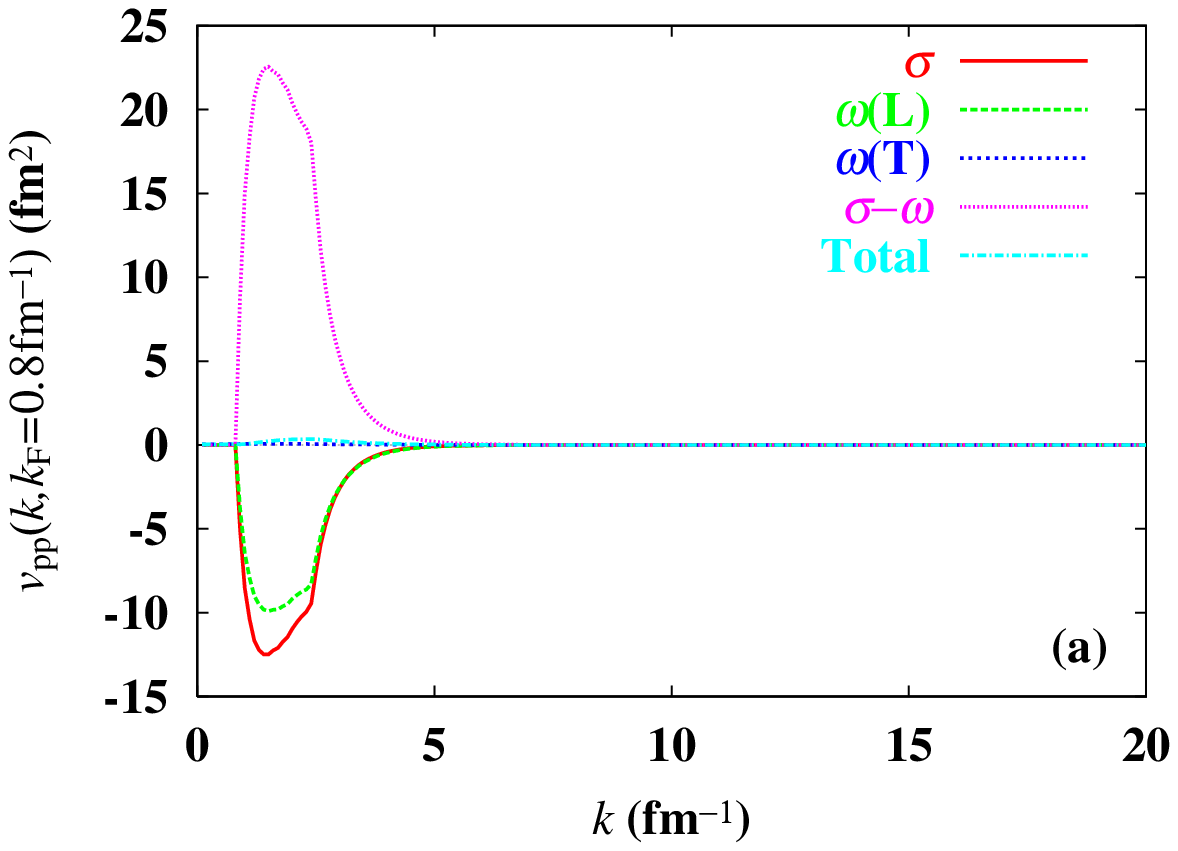}
  \includegraphics[width=7cm]{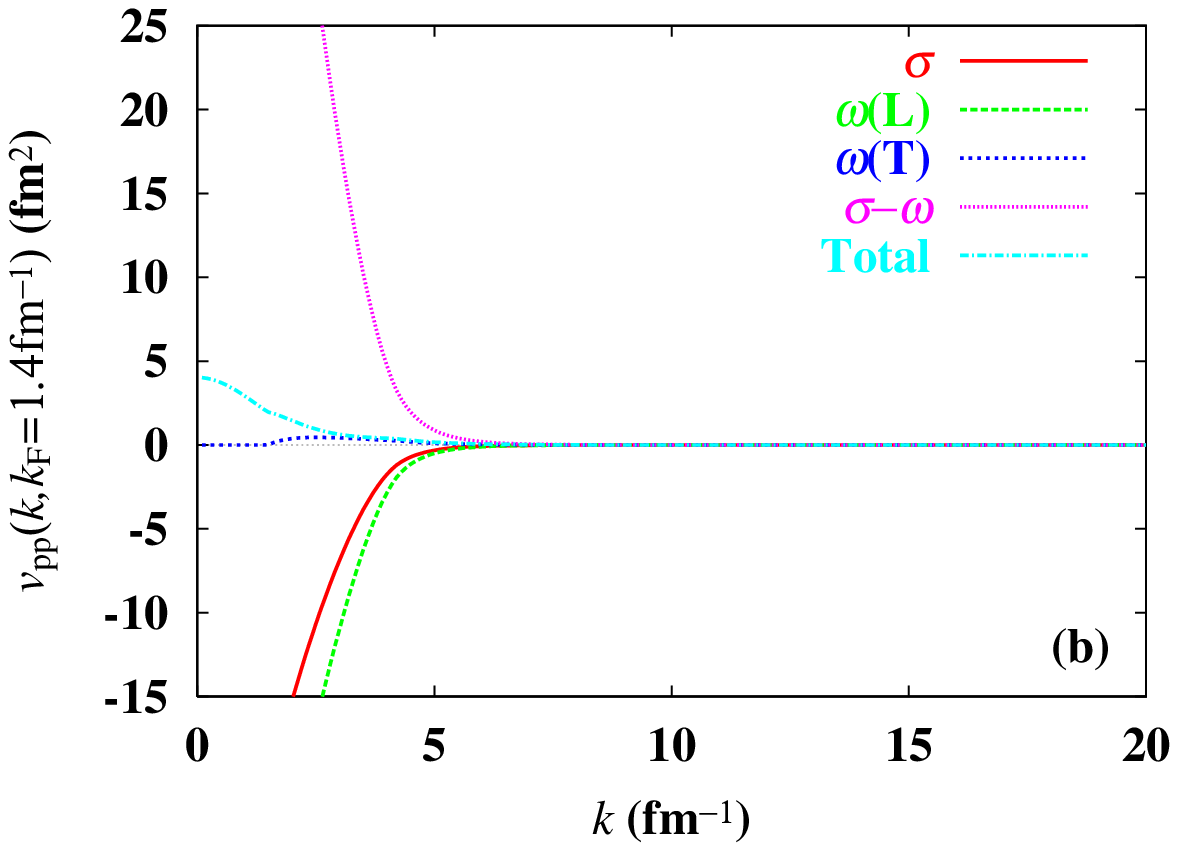}
 \caption{(Color online) The same as Fig.~\ref{fig7} but for the pha case. 
\label{fig8}}
\end{figure}

  Figure~\ref{fig6} also reports the pha case.  The results of the FD and the 
pha cases are very close to each other at low densities.  This is because the 
particle propagation contained $G_\mathrm{D}$ is small.  However, their 
difference grows as density increases;  at high $k_\mathrm{F}$ the polarization 
reduces $\Delta(k_\mathrm{F})$ in contrast to the FD case.  This is brought 
about by the behavior that the total polarization interaction becomes repulsive 
(Fig.~\ref{fig8}).  

\begin{figure}[htbp]
  \includegraphics[width=8cm]{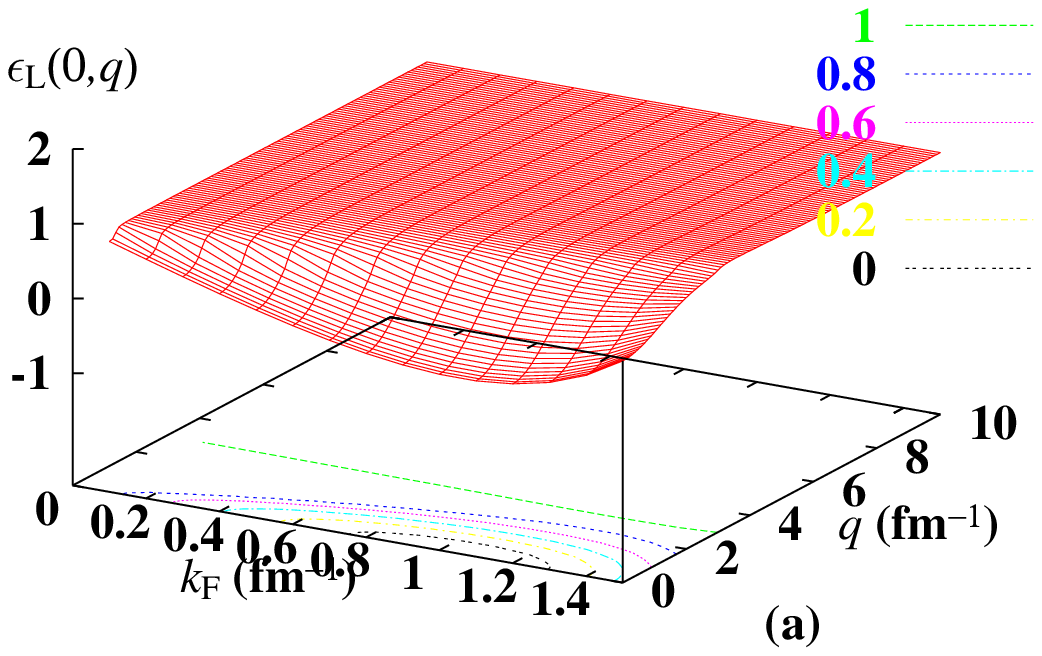}
  \includegraphics[width=8cm]{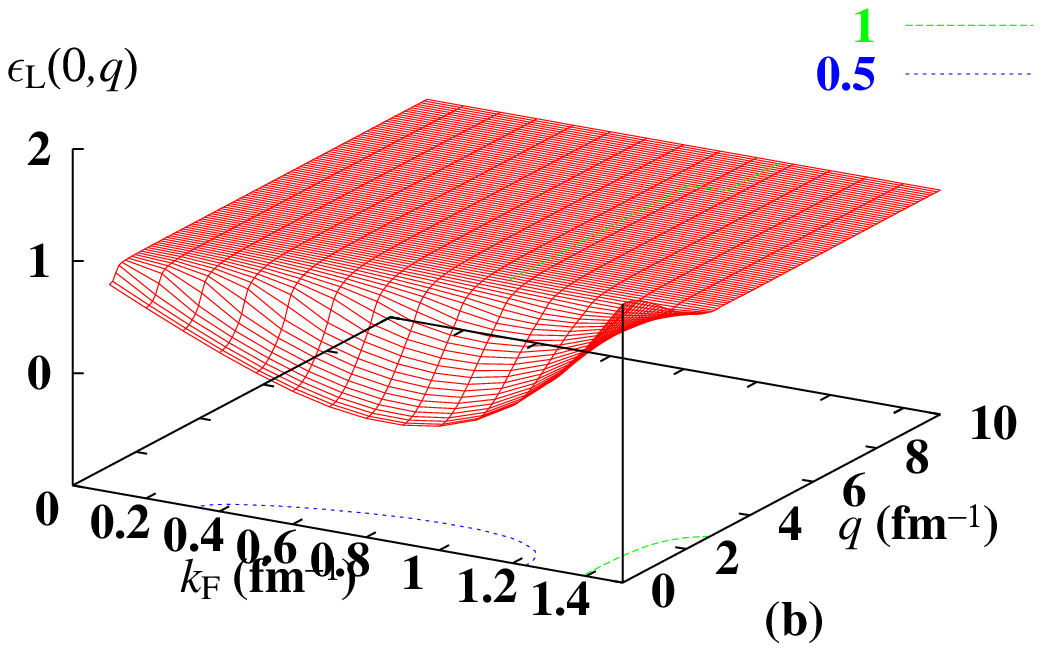}
 \caption{(Color online) The same as Fig.~\ref{fig3} but for pure neutron matter. 
\label{fig9}}
\end{figure}

\begin{figure}[htbp]
  \includegraphics[width=7cm]{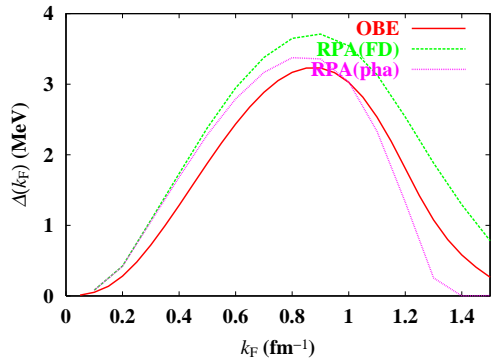}
 \caption{(Color online) The same as Fig.~\ref{fig6} but for pure neutron matter. 
\label{fig10}}
\end{figure}

  Finally we briefly mention the pure neutron matter case.  It is subtle 
whether liquid-gas instability occurs in pure neutron matter or not.  
In the present calculations it occurs in the FD case (Fig.~\ref{fig9}(a)). 
In the pha case it does not occur but $\epsilon_\mathrm{L}$ strongly 
decreases at medium $k_\mathrm{F}$ (Fig.~\ref{fig9}(b)).  Consequently the 
behavior of $\Delta(k_\mathrm{F})$ is similar to the symmetric matter case 
(Fig.~\ref{fig10}).  

\section{Discussion and conclusion}

  In symmetric nuclear matter, the RPA leads to liquid-gas instability at 
moderate densities even at zero temperature.  In the present study, we 
concentrated on the medium polarization, but Ref.~\cite{FH1} showed that 
liquid-gas instability survives even after inclusion of the vacuum 
polarization.  This indicates that the medium polarization effects on 
superfluidity should be considered in the liquid-gas coexistent phase 
rather than in the liquid phase.  In the present investigation, however, 
we have studied superfluidity in the liquid phase, as usual.  Therefore 
only qualitative discussion is allowed for the density region in which 
the instability occurs;  the gap increases from the OBE result.  
In other regions free from the instability, a quantitative discussion is 
possible;  the gap increases in the FD decomposition whereas decreases in 
the pha decomposition at high densities. At low densities the gap increases 
in both cases.  These results are brought about by the 
characteristic cancellation between the attraction from the $\sigma$ 
polarization and the longitudinal $\omega$ polarization and the repulsion 
from the $\sigma$-$\omega$ mixed polarization.  The result for the FD 
case is consistent with that of another relativistic study in 
Ref.~\cite{BCF} and of a non-relativistic study in Ref.~\cite{SLS}.  
Both cases are consistent with Ref.~\cite{He}. 
It is known that the coupling to surface vibrations enhances pairing in 
finite nuclei~\cite{Ba2,Te,Go}.  This tendency is in the same direction as 
the FD case in the present calculation.  But, since the polarization in the 
matter case is density modes, their mutual correspondence should be clarified. 

  It is a subtle problem whether liquid-gas instability occurs in pure 
neutron matter or not.  In the present calculation, it occurs in the FD case.  
Although it does not occur in the pha case, the longitudinal dielectric 
function decreases strongly.  Consequently the results for the gap are 
similar to the symmetric matter case.  Note that the decrease of the Landau 
parameter $F_0$ to around $-1$ was reported in Ref.~\cite{SLS}.  In most of 
non-relativistic calculations~\cite{Cl,Ch2,AWP,WAP,Sc,He,SPR,Sh,SFB}, decrease of 
the gap was reported.  This is due to non-occurrence of liquid-gas instability 
and to the repulsive effect of the spin density fluctuation.  In the present 
model, the effect of 
the transverse $\omega$ polarization that corresponds to the spin density 
mode is weak.  It is worth noting here that a recent ab initio 
calculation~\cite{Fa} concluded that the polarization effect in pure neutron 
matter is weak. 

  It should be stressed that it is an approximation to consider only the 
medium (particle-hole) polarization in the relativistic model;  the vacuum 
(particle-antiparticle) polarization should also be included.  Since the gap 
is sensitive to the tiny remnant of the strong cancellation mentioned above, 
future inclusion of the vacuum polarization would be important.  
This will be studied separately. According to 
Refs.~\cite{JPW,HSK}, the vacuum polarization leads to vector meson mass 
decrease.  In a previous paper~\cite{MT2} we examined this meson mass decrease 
using the in-medium Bonn potential and concluded that it reduces the gap. 

  Other ingredients that are not considered in the present study are inclusion 
of exchange of other mesons like $\pi$ and $\rho$, and the selfenergy 
effects~\cite{Bo,BG,LSZ}.  Study of superfluidity in the liquid-gas coexistent 
phase is also an interesting issue. 

\begin{acknowledgments}
  The author acknowledges Professor P. Ring for suggesting the problem and 
Professor H. Kouno for useful discussions.
\end{acknowledgments}

\appendix

\section{Medium polarization insertions}

\subsection{Feynman-density decomposition}

\begin{gather}
\Pi^\mathrm{S}=\frac{2\lambda g_\sigma^2}{(2\pi)^2}
\Big[k_\mathrm{F}E_\mathrm{F}^\ast-\frac{1}{2}(6M^{\ast\,2}+|\mathbf{q}|^2)
\ln{\Big(\frac{k_\mathrm{F}+E_\mathrm{F}^\ast}{M^\ast}\Big)} \notag \\
+\frac{4M^{\ast\,2}+|\mathbf{q}|^2}{4|\mathbf{q}|}
\Big(2E_\mathrm{F}^\ast-\sqrt{4M^{\ast\,2}+|\mathbf{q}|^2}\Big)
\ln{\Big|\frac{|\mathbf{q}|-2k_\mathrm{F}}{|\mathbf{q}|+2k_\mathrm{F}}\Big|} \notag \\
+\frac{(4M^{\ast\,2}+|\mathbf{q}|^2)^{3/2}}{4|\mathbf{q}|}
\ln{\Big|\frac{E_\mathrm{F}^\ast\sqrt{4M^{\ast\,2}+|\mathbf{q}|^2}+2M^{\ast\,2}+|\mathbf{q}|k_\mathrm{F}}
          {E_\mathrm{F}^\ast\sqrt{4M^{\ast\,2}+|\mathbf{q}|^2}+2M^{\ast\,2}-|\mathbf{q}|k_\mathrm{F}}\Big|}
\Big]\, .
\end{gather}

\begin{gather}
\Pi^\mathrm{L}=-\frac{2\lambda g_\omega^2}{(2\pi)^2}
\Big[\frac{4}{3}k_\mathrm{F}E_\mathrm{F}^\ast-\frac{1}{3}|\mathbf{q}|^2
\ln{\Big(\frac{k_\mathrm{F}+E_\mathrm{F}^\ast}{M^\ast}\Big)} \notag \\
+\frac{1}{6|\mathbf{q}|}
\Big(E_\mathrm{F}^\ast(3|\mathbf{q}|^2-4E_\mathrm{F}^{\ast\,2})
     +(2M^{\ast\,2}-|\mathbf{q}|^2)\sqrt{4M^{\ast\,2}+|\mathbf{q}|^2}\Big)
\ln{\Big|\frac{|\mathbf{q}|-2k_\mathrm{F}}{|\mathbf{q}|+2k_\mathrm{F}}\Big|} \notag \\
-\frac{1}{6|\mathbf{q}|}(2M^{\ast\,2}-|\mathbf{q}|^2)\sqrt{4M^{\ast\,2}+|\mathbf{q}|^2}
\ln{\Big|\frac{E_\mathrm{F}^\ast\sqrt{4M^{\ast\,2}+|\mathbf{q}|^2}+2M^{\ast\,2}+|\mathbf{q}|k_\mathrm{F}}
          {E_\mathrm{F}^\ast\sqrt{4M^{\ast\,2}+|\mathbf{q}|^2}+2M^{\ast\,2}-|\mathbf{q}|k_\mathrm{F}}\Big|}
\Big]\, .
\end{gather}

\begin{gather}
\Pi^\mathrm{T}=-\frac{2\lambda g_\omega^2}{(2\pi)^2}
\Big[-\frac{1}{3}k_\mathrm{F}E_\mathrm{F}^\ast+\frac{1}{3}|\mathbf{q}|^2
\ln{\Big(\frac{k_\mathrm{F}+E_\mathrm{F}^\ast}{M^\ast}\Big)} \notag \\
-\frac{1}{6|\mathbf{q}|}
\Big(E_\mathrm{F}^\ast(2E_\mathrm{F}^{\ast\,2}-6M^{\ast\,2}+\frac{3}{2}|\mathbf{q}|^2)
     +(2M^{\ast\,2}-|\mathbf{q}|^2)\sqrt{4M^{\ast\,2}+|\mathbf{q}|^2}\Big)
\ln{\Big|\frac{|\mathbf{q}|-2k_\mathrm{F}}{|\mathbf{q}|+2k_\mathrm{F}}\Big|} \notag \\
+\frac{1}{6|\mathbf{q}|}(2M^{\ast\,2}-|\mathbf{q}|^2)\sqrt{4M^{\ast\,2}+|\mathbf{q}|^2}
\ln{\Big|\frac{E_\mathrm{F}^\ast\sqrt{4M^{\ast\,2}+|\mathbf{q}|^2}+2M^{\ast\,2}+|\mathbf{q}|k_\mathrm{F}}
          {E_\mathrm{F}^\ast\sqrt{4M^{\ast\,2}+|\mathbf{q}|^2}+2M^{\ast\,2}-|\mathbf{q}|k_\mathrm{F}}\Big|}
\Big]\, .
\end{gather}

\begin{gather}
\Pi^0=\frac{2\lambda g_\sigma g_\omega}{(2\pi)^2}M^\ast
\Big[k_\mathrm{F}+\frac{|\mathbf{q}|^2-4k_\mathrm{F}^2}{4|\mathbf{q}|}
\ln{\Big|\frac{|\mathbf{q}|-2k_\mathrm{F}}{|\mathbf{q}|+2k_\mathrm{F}}\Big|}
\Big]\, .
\end{gather}

\subsection{Particle-hole-antiparticle decomposition}

\begin{gather}
\Pi^\mathrm{S}=\frac{2\lambda g_\sigma^2}{(2\pi)^2}
\Big[\frac{1}{2}k_\mathrm{F}E_\mathrm{F}^\ast-\frac{1}{4}(6M^{\ast\,2}+|\mathbf{q}|^2)
\ln{\Big(\frac{k_\mathrm{F}+E_\mathrm{F}^\ast}{M^\ast}\Big)} \notag \\
-\frac{(4M^{\ast\,2}+|\mathbf{q}|^2)^{3/2}}{8|\mathbf{q}|}
\Big(2\ln{\Big|\frac{|\mathbf{q}|-2k_\mathrm{F}}{|\mathbf{q}|+2k_\mathrm{F}}\Big|}
     -\ln{\Big|\frac{E_\mathrm{F}^\ast\sqrt{4M^{\ast\,2}+|\mathbf{q}|^2}+2M^{\ast\,2}+|\mathbf{q}|k_\mathrm{F}}
          {E_\mathrm{F}^\ast\sqrt{4M^{\ast\,2}+|\mathbf{q}|^2}+2M^{\ast\,2}-|\mathbf{q}|k_\mathrm{F}}\Big|} \notag \\
     +\ln{\Big|\frac{\sqrt{(k_\mathrm{F}+|\mathbf{q}|)^2+M^{\ast\,2}}
                     \sqrt{4M^{\ast\,2}+|\mathbf{q}|^2}+2M^{\ast\,2}+|\mathbf{q}|^2+|\mathbf{q}|k_\mathrm{F}}
                    {\sqrt{(k_\mathrm{F}-|\mathbf{q}|)^2+M^{\ast\,2}}
                     \sqrt{4M^{\ast\,2}+|\mathbf{q}|^2}+2M^{\ast\,2}+|\mathbf{q}|^2-|\mathbf{q}|k_\mathrm{F}}\Big|}
\Big) \notag \\
-\frac{1}{6|\mathbf{q}|}\Big( \big((k_\mathrm{F}+|\mathbf{q}|)^2+M^{\ast\,2}\big)^{3/2}
                             -\big((k_\mathrm{F}-|\mathbf{q}|)^2+M^{\ast\,2}\big)^{3/2} \Big) \notag \\
+\frac{1}{4}k_\mathrm{F}\Big( \sqrt{(k_\mathrm{F}+|\mathbf{q}|)^2+M^{\ast\,2}}
                             +\sqrt{(k_\mathrm{F}-|\mathbf{q}|)^2+M^{\ast\,2}} \Big) \notag \\
-\frac{M^{\ast\,2}}{|\mathbf{q}|}\Big( \sqrt{(k_\mathrm{F}+|\mathbf{q}|)^2+M^{\ast\,2}}
                                      -\sqrt{(k_\mathrm{F}-|\mathbf{q}|)^2+M^{\ast\,2}} \Big) \notag \\
-\frac{E_\mathrm{F}^\ast}{2|\mathbf{q}|}(4M^{\ast\,2}+|\mathbf{q}|^2)
     \ln{\Big|\frac{\sqrt{(k_\mathrm{F}+|\mathbf{q}|)^2+M^{\ast\,2}}-E_\mathrm{F}^\ast}
                   {\sqrt{(k_\mathrm{F}-|\mathbf{q}|)^2+M^{\ast\,2}}-E_\mathrm{F}^\ast}\Big|} \notag \\
+\frac{1}{8}(6M^{\ast\,2}+|\mathbf{q}|^2)
     \ln{\Big|\frac{\big(\sqrt{(k_\mathrm{F}+|\mathbf{q}|)^2+M^{\ast\,2}}+k_\mathrm{F}+|\mathbf{q}|\big)
                    \big(\sqrt{(k_\mathrm{F}-|\mathbf{q}|)^2+M^{\ast\,2}}+k_\mathrm{F}-|\mathbf{q}|\big)}
                    {M^{\ast\,2}}\Big|}
\Big]\, .
\end{gather}

\begin{gather}
\Pi^\mathrm{L}=-\frac{2\lambda g_\omega^2}{(2\pi)^2}
\Big[\frac{2}{3}k_\mathrm{F}E_\mathrm{F}^\ast-\frac{1}{6}|\mathbf{q}|^2
\ln{\Big(\frac{k_\mathrm{F}+E_\mathrm{F}^\ast}{M^\ast}\Big)} \notag \\
+\frac{1}{12|\mathbf{q}|}(2M^{\ast\,2}-|\mathbf{q}|^2)\sqrt{4M^{\ast\,2}+|\mathbf{q}|^2}
\Big(2\ln{\Big|\frac{|\mathbf{q}|-2k_\mathrm{F}}{|\mathbf{q}|+2k_\mathrm{F}}\Big|}
     -\ln{\Big|\frac{E_\mathrm{F}^\ast\sqrt{4M^{\ast\,2}+|\mathbf{q}|^2}+2M^{\ast\,2}+|\mathbf{q}|k_\mathrm{F}}
          {E_\mathrm{F}^\ast\sqrt{4M^{\ast\,2}+|\mathbf{q}|^2}+2M^{\ast\,2}-|\mathbf{q}|k_\mathrm{F}}\Big|} \notag \\
     +\ln{\Big|\frac{\sqrt{(k_\mathrm{F}+|\mathbf{q}|)^2+M^{\ast\,2}}
                     \sqrt{4M^{\ast\,2}+|\mathbf{q}|^2}+2M^{\ast\,2}+|\mathbf{q}|^2+|\mathbf{q}|k_\mathrm{F}}
                    {\sqrt{(k_\mathrm{F}-|\mathbf{q}|)^2+M^{\ast\,2}}
                     \sqrt{4M^{\ast\,2}+|\mathbf{q}|^2}+2M^{\ast\,2}+|\mathbf{q}|^2-|\mathbf{q}|k_\mathrm{F}}\Big|}
\Big) \notag \\
+\frac{1}{15|\mathbf{q}|^3}\Big( \big((k_\mathrm{F}+|\mathbf{q}|)^2+M^{\ast\,2}\big)^{5/2}
                                -\big((k_\mathrm{F}-|\mathbf{q}|)^2+M^{\ast\,2}\big)^{5/2} \Big) \notag \\
-\frac{k_\mathrm{F}}{3|\mathbf{q}|^2}\Big( \big((k_\mathrm{F}+|\mathbf{q}|)^2+M^{\ast\,2}\big)^{3/2}
                                          +\big((k_\mathrm{F}-|\mathbf{q}|)^2+M^{\ast\,2}\big)^{3/2} \Big) \notag \\
+\frac{11}{18|\mathbf{q}|}\Big( \big((k_\mathrm{F}+|\mathbf{q}|)^2+M^{\ast\,2}\big)^{3/2}
                               -\big((k_\mathrm{F}-|\mathbf{q}|)^2+M^{\ast\,2}\big)^{3/2} \Big) \notag \\
-\frac{7}{6}k_\mathrm{F}\Big( \sqrt{(k_\mathrm{F}+|\mathbf{q}|)^2+M^{\ast\,2}}
                             +\sqrt{(k_\mathrm{F}-|\mathbf{q}|)^2+M^{\ast\,2}} \Big) \notag \\
-\frac{2M^{\ast\,2}+5|\mathbf{q}|^2}{6|\mathbf{q}|}\Big( \sqrt{(k_\mathrm{F}+|\mathbf{q}|)^2+M^{\ast\,2}}
                                                        -\sqrt{(k_\mathrm{F}-|\mathbf{q}|)^2+M^{\ast\,2}} \Big)
 \notag \\
+\frac{E_\mathrm{F}^\ast}{6|\mathbf{q}|}(4E_\mathrm{F}^{\ast\,2}-3|\mathbf{q}|^2)
     \ln{\Big|\frac{\sqrt{(k_\mathrm{F}+|\mathbf{q}|)^2+M^{\ast\,2}}-E_\mathrm{F}^\ast}
                   {\sqrt{(k_\mathrm{F}-|\mathbf{q}|)^2+M^{\ast\,2}}-E_\mathrm{F}^\ast}\Big|} \notag \\
+\frac{1}{12}|\mathbf{q}|^2
     \ln{\Big|\frac{\big(\sqrt{(k_\mathrm{F}+|\mathbf{q}|)^2+M^{\ast\,2}}+k_\mathrm{F}+|\mathbf{q}|\big)
                    \big(\sqrt{(k_\mathrm{F}-|\mathbf{q}|)^2+M^{\ast\,2}}+k_\mathrm{F}-|\mathbf{q}|\big)}
                    {M^{\ast\,2}}\Big|}
\Big]\, .
\end{gather}

\begin{gather}
\Pi^\mathrm{T}=-\frac{2\lambda g_\omega^2}{(2\pi)^2}
\Big[-\frac{1}{6}k_\mathrm{F}E_\mathrm{F}^\ast+\frac{1}{6}|\mathbf{q}|^2
\ln{\Big(\frac{k_\mathrm{F}+E_\mathrm{F}^\ast}{M^\ast}\Big)} \notag \\
-\frac{1}{12|\mathbf{q}|}(2M^{\ast\,2}-|\mathbf{q}|^2)\sqrt{4M^{\ast\,2}+|\mathbf{q}|^2}
\Big(2\ln{\Big|\frac{|\mathbf{q}|-2k_\mathrm{F}}{|\mathbf{q}|+2k_\mathrm{F}}\Big|}
     -\ln{\Big|\frac{E_\mathrm{F}^\ast\sqrt{4M^{\ast\,2}+|\mathbf{q}|^2}+2M^{\ast\,2}+|\mathbf{q}|k_\mathrm{F}}
          {E_\mathrm{F}^\ast\sqrt{4M^{\ast\,2}+|\mathbf{q}|^2}+2M^{\ast\,2}-|\mathbf{q}|k_\mathrm{F}}\Big|} \notag \\
     +\ln{\Big|\frac{\sqrt{(k_\mathrm{F}+|\mathbf{q}|)^2+M^{\ast\,2}}
                     \sqrt{4M^{\ast\,2}+|\mathbf{q}|^2}+2M^{\ast\,2}+|\mathbf{q}|^2+|\mathbf{q}|k_\mathrm{F}}
                    {\sqrt{(k_\mathrm{F}-|\mathbf{q}|)^2+M^{\ast\,2}}
                     \sqrt{4M^{\ast\,2}+|\mathbf{q}|^2}+2M^{\ast\,2}+|\mathbf{q}|^2-|\mathbf{q}|k_\mathrm{F}}\Big|}
\Big) \notag \\
+\frac{1}{30|\mathbf{q}|^3}\Big( \big((k_\mathrm{F}+|\mathbf{q}|)^2+M^{\ast\,2}\big)^{5/2}
                                -\big((k_\mathrm{F}-|\mathbf{q}|)^2+M^{\ast\,2}\big)^{5/2} \Big) \notag \\
-\frac{k_\mathrm{F}}{6|\mathbf{q}|^2}\Big( \big((k_\mathrm{F}+|\mathbf{q}|)^2+M^{\ast\,2}\big)^{3/2}
                                          +\big((k_\mathrm{F}-|\mathbf{q}|)^2+M^{\ast\,2}\big)^{3/2} \Big) \notag \\
+\frac{17}{36|\mathbf{q}|}\Big( \big((k_\mathrm{F}+|\mathbf{q}|)^2+M^{\ast\,2}\big)^{3/2}
                               -\big((k_\mathrm{F}-|\mathbf{q}|)^2+M^{\ast\,2}\big)^{3/2} \Big) \notag \\
-\frac{5}{6}k_\mathrm{F}\Big( \sqrt{(k_\mathrm{F}+|\mathbf{q}|)^2+M^{\ast\,2}}
                             +\sqrt{(k_\mathrm{F}-|\mathbf{q}|)^2+M^{\ast\,2}} \Big) \notag \\
-\frac{8M^{\ast\,2}+5|\mathbf{q}|^2}{12|\mathbf{q}|}\Big( \sqrt{(k_\mathrm{F}+|\mathbf{q}|)^2+M^{\ast\,2}}
                                                         -\sqrt{(k_\mathrm{F}-|\mathbf{q}|)^2+M^{\ast\,2}} \Big)
 \notag \\
+\frac{E_\mathrm{F}^\ast}{12|\mathbf{q}|}(4k_\mathrm{F}^2-8M^{\ast\,2}+3|\mathbf{q}|^2)
     \ln{\Big|\frac{\sqrt{(k_\mathrm{F}+|\mathbf{q}|)^2+M^{\ast\,2}}-E_\mathrm{F}^\ast}
                   {\sqrt{(k_\mathrm{F}-|\mathbf{q}|)^2+M^{\ast\,2}}-E_\mathrm{F}^\ast}\Big|} \notag \\
-\frac{1}{12}|\mathbf{q}|^2
     \ln{\Big|\frac{\big(\sqrt{(k_\mathrm{F}+|\mathbf{q}|)^2+M^{\ast\,2}}+k_\mathrm{F}+|\mathbf{q}|\big)
                    \big(\sqrt{(k_\mathrm{F}-|\mathbf{q}|)^2+M^{\ast\,2}}+k_\mathrm{F}-|\mathbf{q}|\big)}
                    {M^{\ast\,2}}\Big|}
\Big]\, .
\end{gather}

\begin{gather}
\Pi^0=\frac{2\lambda g_\sigma g_\omega}{(2\pi)^2}M^\ast
\Big[\frac{1}{2}k_\mathrm{F}-\frac{|\mathbf{q}|^2-4k_\mathrm{F}^2}{4|\mathbf{q}|}
     \ln{\Big|\frac{\sqrt{(k_\mathrm{F}+|\mathbf{q}|)^2+M^{\ast\,2}}-E_\mathrm{F}^\ast}
                   {\sqrt{(k_\mathrm{F}-|\mathbf{q}|)^2+M^{\ast\,2}}-E_\mathrm{F}^\ast}\Big|} \notag \\
+\frac{3E_\mathrm{F}^\ast}{4|\mathbf{q}|}\Big( \sqrt{(k_\mathrm{F}+|\mathbf{q}|)^2+M^{\ast\,2}}
                                         -\sqrt{(k_\mathrm{F}-|\mathbf{q}|)^2+M^{\ast\,2}} \Big) \notag \\
+\frac{M^{\ast\,2}}{2|\mathbf{q}|}
\ln{\Big|\frac{\big(\sqrt{(k_\mathrm{F}+|\mathbf{q}|)^2+M^{\ast\,2}}-E_\mathrm{F}^\ast-|\mathbf{q}|\big)
               \big(\sqrt{(k_\mathrm{F}+|\mathbf{q}|)^2+M^{\ast\,2}}-E_\mathrm{F}^\ast+|\mathbf{q}|\big)}
              {\big(\sqrt{(k_\mathrm{F}-|\mathbf{q}|)^2+M^{\ast\,2}}-E_\mathrm{F}^\ast-|\mathbf{q}|\big)
               \big(\sqrt{(k_\mathrm{F}-|\mathbf{q}|)^2+M^{\ast\,2}}-E_\mathrm{F}^\ast+|\mathbf{q}|\big)}\Big|}
\Big]\, .
\end{gather}

\end{document}